\documentstyle[preprint,aps]{revtex}

\tightenlines

\newcommand{\lsim}[1]{
\setlength{\unitlength}{12pt}
\begin{picture}(1.4,1.)
\put(.7,-0.3){\makebox(0.0,1.)[t]{$<$}}
\put(.7,-0.3){\makebox(0.0,1.)[b]{$\sim$}}
\end{picture}#1}
\newcommand{\gsim}[2]{
\setlength{\unitlength}{12pt}
\begin{picture}(1.4,1.)
\put(.7,-0.3){\makebox(0.0,1.)[t]{$>$}}
\put(.7,-0.3){\makebox(0.0,1.)[b]{$\sim$}}
\end{picture}#2}
\begin{document}
\draft

\title{Resonant conversions of extremely high energy neutrinos in dark
matter halos}

\author{R. Horvat \\
``Rudjer Bo\v skovi\' c'' Institute, P.O.Box 1016, 10001 Zagreb,
Croatia}

\maketitle

\begin{abstract}

We study the effect of adiabatically resonant conversion in galactic halos
of neutrinos at the highest energies ($ \sim 10^{20}$ - $10^{22}$ eV), when
the $\nu$ source is in the center of a galaxy. Using the standard neutrino
properties and the standard cosmological scenario for the hot dark part of 
matter, we find that interesting conversions may take place just for
neutrino parameters relevant to the solar and atmospheric neutrino problem.
The effect is due to the large enhancement in the $\nu$ density in galactic
halos and to the form of the effective matter potential both below and above
the pole of the $Z$ resonance.

\end{abstract}

%\pacs{14.60Pq, 95.85Ry, 95.35+d, 98.70Sa}
\newpage

It has recently been suggested \cite{1,2,3} that annihilation of extremely
high energy (EHE) ($ \gsim \;\mbox{\rm 10}^{21} \;\mbox{\rm eV}$) primary cosmic
neutrinos on a background of clustered relic neutrinos may be able to
account for the detected flux of cosmic rays of the highest energy ($ \gsim
\;\mbox{\rm 10}^{20} \;\mbox{\rm eV}$) \cite{4,5}. The Fly's Eye \cite{4}
and AGASA \cite{5} experiments confirmed the existence of proton-like cosmic
rays with energies $\gsim \;\mbox{\rm 10}^{20} \;\mbox{\rm eV}$ which, being 
well above Greisen-Zatsepin-Kuzmin \cite{6} cut-off, may create a rather 
serious problem in terms of a possible astrophysical explanation. In the above
scenario \cite{1,2,3} this problem is elegantly circumvented by the fact
that the highest energy cosmic rays may actually be generated very close to
Earth, possibly in our hot-dark-matter galactic halo \cite{1,2}. In order
that this new mechanism for generating the highest energy cosmic rays has a
chance to work, a relic electron-volt mass neutrino would be needed as the
gravitational clustering of relic cosmic neutrinos is more efficient for
heavier and slower species. If the Hubble and age parameter allow for a
critical-density universe \cite{7} as suggested by inflation, then the mixed
Cold + Hot Dark Matter model with $m_{\nu_{\mu}} \simeq m_{\nu_{\tau}}
\simeq 2.5 \;\mbox{\rm eV}$ (giving $\Omega_{\nu} \simeq 0.2$) was found to be
consistent with all available observations. Indeed this model was shown to
agree well with observations of cosmic background temperature anisotropy on
large scales performed by the COBE satellite \cite{8} as well as with those on
the number density of clusters \cite{9}. Although in accordance with the
theory of formation of cosmic structure, the scenario is beset by the
problem \cite{3} that one requires a very large energy generation rate of
high energy neutrinos (which poses a difficult astrophysical problem) and
the fact that $Z, W^{\pm}$ decay produces a large population of $\pi^{0}$
secondaries and hence we have more photons (from $\pi^{0}$ decay) than
protons. 

The interaction of the EHE neutrino with the relic one had been studied
first in \cite{10} and reconsidered afterwards in \cite{11}. 
It was shown \cite{11} that neutrinos with the standard
electro-weak interaction can travel through cosmic distances with a little
hindrance, and only those neutrinos with energies close to the $Z$
resonance value may be absorbed while crossing galactic halos, leading to a
dip in the spectrum around $E_{res} \simeq \;\mbox{\rm 10}^{21} \;\mbox{\rm
eV}$. This idea was then taken over in \cite{1,2} to show that secondaries
(such as protons and photons) of the EHE-relic neutrino annihilation in our 
galactic halo may be responsible for the observed high energy cosmic ray energy
spectrum.

Here we examine whether the condition for adiabatically resonant conversion
of these EHE neutrinos  while traversing the halo around the source is met, if
the oscillation parameters are taken to be just those responsible for the
solution to the solar and atmospheric neutrino anomaly. In the following we
will be considering $\nu_e -\nu_{\tau(\mu)}$, $\nu_e -\nu_s$ ($\nu_s$ being a
sterile neutrino) as well as $\nu_{\mu} -\nu_s$ oscillations.   
 
Assuming that the halo is nearly round we take for the halo density a
density profile like 
\begin{equation}
\rho (r) = \frac{\rho (r=0)}{1 + (r/a)^{2}} \; ,
\label{form1}
\end{equation}
where $a$ is the core radius of the halo. By considering the values for our 
local galactic halo it was found \cite{1} that, 
notwithstanding the fact that light neutrinos
constitute a smaller fraction of the density of the galactic halo \cite{12}
then they do of the cosmological density in a typical mixed-dark-matter
scenario \cite{13}, the neutrino number density inside the core might be a
factor $ \sim \;\mbox{\rm 10}^{5}-\;\mbox{\rm 10}^{7}$ larger than the average
cosmic $\nu$ density. Hence, $ N_{\nu} \simeq
\;\mbox{\rm 10}^{7}-\;\mbox{\rm 10}^{9} \; \mbox{\rm cm}^{-3}$ was assumed 
in \cite{1}, together with a relic neutrino mass $m_{\nu}=10 \;\mbox{\rm
eV}$. In the present work, however, we treat $N_{\nu}$ as a free parameter,
presenting first the conditions under which conversion may occur, and then
adding a discussion of whether or not these conditions may be met in
`realistic neutrino halo' allowed by the Tremaine-Gunn phase-space
constraint \cite{14}.
In addition, we assume that the halo consists predominantly of heavier (
$\nu_{\mu}$ and $\nu_{\tau}$) neutrinos with $m_{\nu_{\mu}} \simeq
m_{\nu_{\tau}} \simeq 2.5 \;\mbox{\rm eV}$. It is just the fact that the
background is {\sl not} flavor symmetric that allows us to study not
only active-sterile but  active-active neutrino oscillations as well.

Effects of a medium on neutrino propagation is determined by the difference
of potentials, whose standard-model contribution in the context of Thermal
Field Theory may  readily be obtained from the relevant thermal
self-energies of a neutrino: charged-current, neutral-current and tadpole.
Of these, only  neutral-current and tadpole contribution are of relevance
here as the halo background consists of pure neutrinos (up to some of
whatever dark-matter particles which appear as a major constituent of the
halo and which are irrelevant for the present discussion). The contribution
from the tadpole diagram, beside of being a constant independent of the
external neutrino momentum and hence irrelevant (see below), is always
proportional to the asymmetry for certain neutrino species. Since the cosmic 
 neutrino background is likely to be $CP$-symmetric its contribution is
completely negligible, leaving us with the neutral-current contribution
only.

Since the `target' neutrinos are massive ($m_{\nu} \simeq 2.5
\;\mbox{\rm eV}$) the $Z$ boson may be considered ``massless'' always when $
s \simeq 2E_{\nu}m_{\nu} \gg M_{Z}^{2}$. This corresponds to
energies in the lab frame $E_{\nu} \gsim \;\mbox{\rm 10}^{21} \;\mbox{\rm eV}$. In the
opposite limit, $s \ll M_{Z}^{2}$, the $Z$ boson should be considered
``massive'' and the usual contact approximation for the $Z$-propagator is
adequate. Using the real-time version of Thermal Field Theory, we find by
explicit calculation a contribution to the effective matter potentials in
the ``massless'' limit as
\begin{equation}
V_e \simeq 0.02 \; (T_{\nu}^{2}/E_{\nu})\; (\sum_{i=2,3} |U_{ei}|^{2}
 )\;\;;
\label{form2}
\end{equation}
\begin{equation}
V_{\mu} \simeq V_{\tau} \simeq 0.02 \; (T_{\nu}^{2}/E_{\nu})\;.
\label{form3}
\end{equation}
Although the heaviest neutrinos are so massive that they are firmly
non relativistic today they still should be described by the relativistic
distribution function as their total number density is fixed at 
about $\mbox{\rm 100} \; \mbox{\rm cm}^{-3}$ in the uniform, non-clustered
background and at about a factor of few larger in the galactic halo. 
Hence the temperature $T_{\nu}$
from Eqs.(2) and (3) is always related to the neutrino number density via
the relation, $T_{\nu} \simeq (N_{\nu}/0.18)^{1/3}$. Notice, in addition,
that the appearance of a unitary mixing matrix in the vacuum $U$ in the
expression for $V_e$ in (2) is due to the fact that internal lines of the
thermal self-energy graph should necessarily be written in the (vacuum)
mass eigenstate basis. If in a world with just three neutrinos all of them
have nearly the same mass, of about $ \sim \; \mbox{\rm 1.5}\;\mbox{\rm
eV}$, then the suppression due to ($\sum_{i} |U_{ei}|^{2}$) will be absent.  
 
Let us first consider the values of the parameters of the small mixing angle
$\nu_{e}-\nu_s$ solution to the solar neutrino problem which lie in the
region \cite{15}:
\begin{equation}
2.9 \times 10^{-6} \; \lsim \Delta m_{es}^{2}/eV^{2} \; \lsim 7.7 \times
10^{-6}  \; \; ;\; \;
3.5 \times 10^{-3} \; \lsim \sin^{2}{2 \theta} \; \lsim 1.4 \times 10^{-2}
\;,
\label{form4}
\end{equation}
where $\Delta m_{es}^{2}=m_s^{2}-m_e^{2}$ and $\theta$ is the vacuum mixing
angle of the $\nu_e -\nu_s$ mixing system.
The MSW resonant condition for
all neutrinos with $E_{\nu} > \mbox{\rm 10}^{21} \;\mbox{\rm
eV}$ now reads
\begin{equation}
0.04 \; T_{\nu}^{2} \; (\sum_{i} |U_{ei}|^{2})\; \simeq \Delta m_{es}^{2} \;.
\label{form5}
\end{equation}
Note that Eq.(5) is independent of neutrino energy. For the central value ($
5 \times \mbox{\rm 10}^{-6}$) in (4) we find that the MSW
resonance for $\nu_e$ -$\nu_s$ oscillations occurs within the core radius of
the halo if the neutrino number density inside the core of a galaxy is $
 3 \times {\mbox{\rm 10}^{7}} (\sum_{i} |U_{ei}|^{2})^{-3/2}\;\mbox{\rm
 cm}^{-3}$.
The same occurs also for ${\bar{\nu}}_e \rightarrow
{\bar{\nu}}_{s}$ oscillations as $V_{\bar{e}}>0$.

The condition for unsuppressed oscillations is that the oscillation frequency
must be real; otherwise the system is critically overdamped, the
oscillations would be fully incoherent, and hence in fact there will be no
oscillations. This translates into the requirement that the oscillation
frequency should be much greater than the total neutrino interaction rate
$\Gamma$,
\begin{equation}
2\pi \omega_{osc} \; > \; \frac{1}{2} \Gamma\;. 
\label{form6}
\end{equation}
At the resonance this gives
\begin{equation}
\frac{\Delta m^{2} \; \sin 2\theta }{2E_{\nu}}\;>\;\frac{1}{2} \Gamma\;.
\label{form7}
\end{equation}
For $E_{\nu} = \mbox{\rm 10}^{22}\;\mbox{\rm eV}$ we have $(\Delta m^{2}
\sin 2\theta )/2E_{\nu} \; \simeq \; 2 \times \mbox{\rm 10}^{-29}\;
\mbox{\rm eV}$, whereas $(1/2) \Gamma = (1/2)N_{\nu} \sigma \; \simeq 
\; 5 \times 
(\mbox{\rm 10}^{-32}- \mbox{\rm 10}^{-30})\; \mbox{\rm eV}$ for $ N_{\nu}
\simeq
\;\mbox{\rm 10}^{7}-\;\mbox{\rm 10}^{9} \; \mbox{\rm cm}^{-3}$. 
The latter
estimate follows from \cite{11}, where the total neutrino absorption cross
section was calculated to be $\sigma (E_{\nu} = \mbox{\rm
10}^{22}\;\mbox{\rm eV})\; \simeq \; 5 \times \mbox{\rm 10}^{-34}\;
\mbox{\rm cm}^{2}$. Thus $\nu_e -\nu_s$ oscillations in the halo of a galaxy
can be considered as unsuppressed for $E_{\nu}\; \sim \; \mbox{\rm
10}^{22}\; \mbox{\rm eV}$. However, level crossing alone is not sufficient
for a complete conversion. The extent of conversion is determined by a
degree of adiabacity; if there is only level crossing but no sufficient
adiabacity at the resonance then there will be no complete conversion of
electron neutrinos into sterile neutrinos. The condition for adiabatic
transition at the resonance may be written in our case as     
\begin{equation}
\frac{3 \Delta m^{2} \sin^{2}{2 \theta }}{4 E_{\nu} \cos 2\theta 
\mid d N_{\nu}/N_{\nu} dr \mid_{0}}\; \gg \; 1\;,
\label{form8}
\end{equation}
where the subscript $0$ denotes that this quantity should be evaluated at
the resonance. Using the distribution (1) one can easily show
that for $E_{\nu}\; \sim \; \mbox{\rm 10}^{22}\;\mbox{\rm eV}$ adiabatically
resonant conversion may only take place  in the regions very close to the
center of a galaxy, namely for 
\begin{equation}
r_0 \; \lsim \; 2 \times 10^{-3} \; a \; \simeq \cal O  (\mbox{\rm 10} 
\; \mbox{\rm  pc}) \;.
\label{form9}
\end{equation}
For the last estimate in (9) we have taken 
$a \simeq \mbox{\rm 10}\; \mbox{\rm kpc}$
\cite {16}. Eq.(9) is due to the fact that the adiabacity parameter, as
defined in (8), can be larger than unity for such EHEs only where the
flatness of the distribution (1) is appreciable. There is also another
solution for $r_0$ from (8) but it is located well outside the core where
the number density is smaller; besides, Eq.(1) is not a good fit to $\rho
(r)$ at $r \gg a$ \cite {17}. However even the restriction (9) is consistent
with the models \cite{18,19} in which neutrino production takes place in 
Active Galactic Nuclei (AGN)
jets. In these models protons are accelerated in the jets that stream highly
collimated out of their cores to distances of several parsec. In the jets
the neutrinos are Lorentz boosted to energies much higher than in AGN cores,
what has important implications for their detection. Notice, for comparison,
that $V_e$ obtained here is always larger (for $(\sum_{i}
|U_{ei}|^{2})^{3/2}\;\gsim \; \mbox{\rm 10}^{-2}$ and $ N_{\nu}\;\simeq
\;\mbox{\rm 10}^{7}\; \mbox{\rm cm}^{-3}$) 
than the
effective matter potential in AGN cores whose magnitude was estimated to be 
$\mbox{\rm 10}^{-29}-\mbox{\rm 10}^{-32}\; \mbox{\rm eV}$ \cite{20}. One should
stress however here that as $\rho (r)$ has a flat region for $r < r_0 $, the
neutrinos are always produced at densities very close to the resonant one.
Thus, the mixing angle at the production point $\theta_{mp} \; \simeq \;\pi
/4$ rather than $\theta_{mp} \rightarrow \pi /2$, and consequently the
oscillation probability $P(\nu_e \rightarrow \nu_e )$ is centered around
$1/2$ rather than around $\sin^{2}{\theta}$. One can therefore speak of
partial ($\gsim \mbox{\rm 50}\%$) conversion.   

Let us now turn to active-active $\nu_e \rightarrow \nu_{\tau (\mu )}$
oscillations. Notice that there is no resonant transition in the
``massless'' limit as $\Delta m^{2}_{e \tau (\mu)} > 0$, but $V_e - V_{\tau
(\mu)} < 0$ [see Eqs.(2) and (3)]. The same holds for 
${\bar{\nu}}_e \rightarrow
{\bar{\nu}}_{\tau (\mu )}$ oscillations as 
$V_{\bar{e}} - V_{\bar{\tau} (\bar{\mu})} < 0$.
The situation turns out to be completely different if we concentrate to the
``massive'' limit for the $Z$-boson propagator. In this case the effective
matter potential is given as a sum of the finite-density and
finite-temperature contribution
\begin{equation}
V(E_{\nu})\; = \; V_{\rho}\;+\; V_T(E_{\nu})\;,
\label{form10}
\end{equation}
where [for $\nu_{\tau} (\nu_{\mu})$]
\begin{equation}
V_{\rho}\; \simeq \; \sqrt{2} G_F N_{\gamma} L_{\nu}                                                             
\label{form11}
\end{equation}
and
\begin{equation}
V_T(E_{\nu})\; \simeq \; -24 G^{2}_F T^{4}_{\nu} E_{\nu}\;.                                                              
\label{form12}
\end{equation}
In (11) $N_{\gamma}$ represents the number density of photons and $L_{\nu}$
the tau (muon)-like asymmetry. We are now in position to examine a
possibility for resonant conversion as we have $V_e - V_{\tau (\mu)}\;
\simeq -V_{\tau} > 0$. It is to be noted that for $L_{\nu} = \mbox{\rm
10}^{-10}$ and $E_{\nu}\; \sim \; \mbox{\rm 10}^{20}\; \mbox{\rm eV}$, $\mid
V_T \mid $, which represents a contribution coming from the low-energy tail
of the $Z$ resonance, is about $7$ orders of magnitude larger than
$V_{\rho}$ for $ N_{\nu}\;\simeq \;\mbox{\rm 10}^{7}\; \mbox{\rm cm}^{-3}$
, and therefore from (10) $V \; \simeq V_T \; <\; 0$. 

Let us now concentrate to the hypothesis of two-neutrino vacuum oscillations
which was found to provide a good quality description of the solar neutrino
data for values of the two oscillation parameters belonging approximately to
the region \cite{21}
\begin{equation}
5.0 \times 10^{-11} \; \lsim \Delta m_{e \tau (\mu)}^{2}/eV^{2} \; \lsim 
10^{-10}  \; \; ;\; \;
0.65 \; \lsim \sin^{2}{2 \theta} \; \leq \; 1.0
\;.
\label{form13}
\end{equation}
The resonant energy for the central values in (13) ($7.5 \times \mbox{\rm
10}^{-11},\; \sin^{2}{2 \theta}\; \simeq \; \mbox{\rm 0.8}$) turns out to
lie in the range $\mbox{\rm 10}^{19}-\mbox{\rm 10}^{21}\; \mbox{\rm eV}$ for
the input neutrino number density $\mbox{\rm 10}^{9}-\mbox{\rm 10}^{7}\; 
\mbox{\rm cm}^{-3}$. The same occurs also for ${\bar{\nu}}_e \rightarrow 
{\bar{\nu}}_{\tau (\mu )}$ oscillations as $V_{\bar{e}} - V_{\bar{\tau}
(\bar{\mu})} > 0$. Here we would like to stress the importance of
$\nu_e-\nu_{\tau}$ oscillations as the initial fluxes of the ultra-high
energy neutrinos originating from AGNs are estimated to have a ratio
\cite{22}
$\nu_{\tau}/\nu_{\mu}\; \lsim \; \mbox{\rm 10}^{-3}$ (also 
$\nu_{e}/\nu_{\mu}\; \simeq \; 1/2$ is expected from AGNs). Hence if an 
enhanced $\nu_{\tau}/\nu_{e}$ ratio as compared to the non oscillation case is
observed correlated to the direction of the source for ultra-high energy
neutrinos, then this may represent a piece of evidence for neutrino
oscillations taking place in the halo around the source. The model of
Ref.\cite{18} with maximum energy of $\mbox{\rm 10}^{18}-\mbox{\rm
10}^{19}\; 
\mbox{\rm eV}$ can
therefore be very relevant to our scenario where level crossing occurs at a
bit higher energies. We find that resonant conversion with sufficient
adiabacity does occur in the part of a dark halo with
\begin{equation}
r_0 \; \lsim \;\cal O  (\mbox{\rm few}\;\mbox{\rm  pc}) \;.
\label{form14}
\end{equation}
This is very similar to the estimate (9) for $\nu_e -\nu_s$ oscillations in
the ``massless'' limit and hence still compatible to the proposed sites for
neutrino production in the vicinity of AGNs.
 
Let us finally consider the $\nu_{\mu}-\nu_s$ \cite{23} oscillation
hypothesis that provides an appealing explanation for the large up-down
asymmetry observed by SuperKamiokande for atmospheric neutrino induced
$\mu$-events \cite{24}. The oscillations $\nu_{\mu}-\nu_s$ can solve the 
atmospheric problem if oscillation parameters lie in the approximate range
\begin{equation}
3 \times 10^{-4} \; \lsim \Delta m_{\mu s}^{2}/eV^{2} \; \lsim
7 \times 10^{-3}  \; \; ;\; \;
0.82 \; \lsim \sin^{2}{2 \theta} \; \leq \; 1.0
\;.
\label{form15}
\end{equation}
We note that it has been claimed \cite{25} recently that at present there is
no constraint from cosmology to the $\nu_{\mu}-\nu_s$ solution with the
aforementioned parameters.

Since $\mid 2E_{\nu} V_{\mu} \mid << \mid \Delta m_{\mu s}^{2} \mid $,
level crossing never occurs in a halo expect when $\sin^{2}{2 \theta}$ is
extremely close to the unity. Rather than considering such a coincidence we
check out the adiabatic condition
\begin{equation}
(\sin^{2}{2 \theta})^{-1}\; \frac{( \Delta m_{\mu s}^{2})^{2}}
{2E^{2}_{\nu} \mid V_{\mu} \mid}\;>> \; a^{-1}\;.                                              
\label{form16}
\end{equation}
While in the ``massless'' limit with $E_{\nu} \sim \mbox{\rm 10}^{22}\;
\mbox{\rm eV}$ the condition (16) is well satisfied for the whole range 
$0 \; \lsim \; r \; \lsim a$
, it is always extremely
well satisfied in the ``massive'' limit. It is well known \cite{26} that 
actually  it does not matter whether a  resonance has been crossed or not
while a neutrino propagates adiabatically from one region to another as
the similar formula for the propagation probability applies
\begin{equation}
P(\nu_{\mu} \rightarrow \nu_{\mu})\;=\;\frac{1}{2}(1+\cos{{\theta}_{im}}
\cos{{\theta}_{fm}})\;.                                                 
\label{form17}
\end{equation}                                
Eq.(17) describes a situation where $\nu_{\mu}$ produced at some density
near the galactic center, with mixing angle ${\theta}_{im}$,  
propagates adiabatically through the halo to the region of diffuse neutrino 
background, with mixing angle ${\theta}_{fm}$. Since the mass squared
difference $\Delta m_{\mu s}^{2}$ is always larger than the induced mass
squared of the muon neutrino, the neutrino does not go through the resonance
but is produced below it, and therefore we have
\begin{equation}
{\theta}_{im}\;\simeq \; {\theta}_{fm} \; \simeq \theta \;.
\label{form18}
\end{equation}
In this case the propagation probability (17) reduces just to the classical
probability in the vacuum. With the parameters from (15) we get a constraint
\begin{equation}
0.5\; \lsim \; P(\nu_{\mu} \rightarrow \nu_{\mu})\; \lsim \; 0.59 \;.
\label{form19}                                                   
\end{equation}
   
We have seen that all the above effects except that based on the
$\nu_{\mu}-\nu_s$ oscillation hypothesis require at least $ N_{\nu}\;\simeq
\;\mbox{\rm 10}^{7}\; \mbox{\rm cm}^{-3}$. There is, however, a strong bound
based on Pauli's exclusion principle on the number density $N_{\nu}^{max}$
of neutrinos clustered in the halo around a galaxy. The neutrino density in
the halo is approximately constrained because of this requirement
(Tremaine-Gunn constraint \cite{14}) by 
$N_{\nu}^{max} \leq  (4 \pi /3)(p_{max}/h)^{3}$,
where $p_{max} = m_{\nu} v_{esc}$ is the maximum momentum for a neutrino of
mass $m_{\nu}$ and $v_{esc}$ is the escape velocity of bound neutrinos. If,
in addition, we assume that massive neutrinos that fill up dark matter halos
today result from gravitational instability of three cosmic seas of relic
neutrinos, then the bound is improved by a factor of 2, giving
\begin{equation}
N_{\nu}^{max} < 1 \times 10^{5} \; \left ( \frac{m_{\nu}}{2.5 \; \mbox{\rm eV}} 
\right )^{3}\;\left ( \frac{v_{esc}}{500 \; \mbox{\rm km/s}} \right )^{3} 
\;\mbox{\rm cm}^{-3}\;\;.
\label{form19}
\end{equation}  
It is clear from Eq.(20) that to have sufficiently large $N_{\nu}$ in halos
where the neutrinos have typical escape velocities, larger values of
$m_{\nu}$ would be preferable. Conversely, if we stick to the Cold + Hot
Dark Matter scenario of structure formation ($m_{\nu_{\mu}} \simeq
m_{\nu_{\tau}} \simeq 2.5 \;\mbox{\rm eV}$), the increased neutrino density
can be achievable only in halos where the neutrinos have large escape
velocities. Notice, however, that larger values of $m_{\nu}$ are still
tolerable by the cosmological mass bound on light neutrinos \cite{27}
($m_{\nu_e} + m_{\nu_{\mu}} + m_{\nu_{\tau}} \lsim 37 \; \mbox{\rm eV}$), and
thus for example $m_{\nu_{\mu}} \simeq m_{\nu_{\tau}} \simeq 10
\;\mbox{\rm eV}$ may result in $ N_{\nu}\;\simeq \;\mbox{\rm 10}^{7}\;
\mbox{\rm cm}^{-3}$. The effect of the higher neutrino masses would be to
lower the resonant energy $E_{\nu}^{res}$ (the energy which discriminate 
between the ``massive'' and ``massless'' limit) by a factor of few, thereby
increasing the $\nu_e-\nu_s$ conversion and depressing $\nu_e-\nu_{\tau
(\mu)}$ conversion.  

In conclusion, we have considered a possibility for adiabatically resonant
conversion of EHE neutrinos originating from the center of a galaxy, while
traversing its dark matter halo. Owing to the large enhancement of the
effective neutrino potentials at energies immediately above the pole of the
$Z$ resonance, we have found that $\nu_e$'s can resonantly be converted into
unobservable $\nu_s$'s, if their parameters are consistent with the MSW
interpretation of the solar neutrino data. On the other hand, because the
hot dark matter around galaxies is expected not to be flavor symmetric,
$\nu_e$'s can  resonantly be converted into $\nu_{\tau}$'s, if their
parameters are consistent with the vacuum mixing solution of the solar
neutrino anomaly. This kind of transition is especially important because
there is a negligible number of $\nu_{\tau}$'s in the initial fluxes of high
energy neutrinos of astrophysical origin. Since the resonance density
approaches the central neutrino production region, the above conversions
turn out to be only partial. It is strikingly that in some proton blazar
models a maximum neutrino energy is very close to the resonant energy. With
regard to the Tremaine-Gunn constraint, these effects require either
neutrino masses close to the cosmological mass bound or such halos where the
neutrinos have large escape velocities, or both. 
Finally, if the parameters of active-sterile mixing system are consistent
with
the $\nu_{\mu}-\nu_s$ solution to the atmospheric neutrino anomaly, then
neutrino propagation inside a dark matter halo is always adiabatic for 
$E_{\nu}\; \lsim \; \mbox{\rm 10}^{24}\; \mbox{\rm eV}$ and the propagation
probability simply reduces to the classical probability in the vacuum.

The author acknowledges the support of the Croatian Ministry of Science and
Technology under the contract 1 -- 03 -- 068.

\end{document}